\documentclass[submission,copyright,creativecommons]{eptcs}
\providecommand{\event}{QSQW2020} 
\usepackage{breakurl}             
\usepackage{underscore}           
\usepackage{amsmath}
\usepackage{amsfonts}
\usepackage{amsthm}
\usepackage{dsfont}
\usepackage{graphics}
\usepackage{graphicx}
\usepackage{subfigure}
\usepackage{amssymb}
\usepackage {mathrsfs}
\usepackage{braket}

\title{The Gambler's Ruin Problem and Quantum Measurement}
\author{Fabrice Debbasch
\institute{Sorbonne Universit\'e, Observatoire de Paris, Universit\'e PSL, CNRS, LERMA, F-75005, Paris, France}
\email{fabrice.debbasch$@$gmail.com
}}
\def\titlerunning{The Gambler's Ruin and Quantum Measurement}
\def\authorrunning{F. Debbasch}
\begin{document}
\maketitle
\begin{abstract}
The dynamics of a single microscopic or mesoscopic non quantum system interacting with a macroscopic environment is generally stochastic. In the same way, the reduced density operator of a single quantum system interacting with a macroscopic environment is {\sl a priori} a stochastic variable, and decoherence describes only the average dynamics of this variable, not its fluctuations. It is shown that a general unbiased quantum measurement can be reformulated as a gambler's ruin problem where the game is a martingale. Born's rule then appears as a direct consequence of the optional stopping theorem for martingales. Explicit computations are worked out in detail on a specific simple example.
%
%
%
%
%
\end{abstract}

\section{Introduction}

Quantum physics, as it was elaborated in the first half of the 20th century, needs two main ingredients. One is a description of the unitary time evolution of quantum systems in the absence of measurement (e.g. the Schr\"odinger equation) \cite{Pauli80a} and the other is a rule which predicts the probabilities of various measurements. This rule was introduced by M. Born in 1926 \cite{B26a} and now bears his name.

Understanding if and how the Born rule emerges from unitary evolutions of quantum system has been a longstanding problem of quantum mechanics \cite{Jammer66a, Omnes94a}. Most physicists now agree that decoherence \cite{Joos96a, Schlosshauer07a, BreuerPetr07a} delivers at least part of the answer. But physicists disagree \cite{Schlosshauer07a} on whether decoherence tells the whole story, or if it should be complemented by something else, like for example explicit collapse models.


Decoherence predicts that, upon measurement, the density operator of a quantum macroscopic system interacting with its environment evolves toward a classical superposition of states and that no quantum interference is therefore observed on macroscopic objects. What is still a matter of debate is how the apparent random character of measurement articulates with decoherence. There are essentially two points of view. The first point of view considers that the classical superposition of states predicted by decoherence never collapses on a single state (see for example H.D. Zeh's contribution in \cite{Joos96a}). The apparent collapse is then purely subjective `in the observer's mind'. This point of view seems to warrant a many-world, {\sl a la} Everett- DeWitt interpretation of quantum 
physics \cite{Dewiit16a}. 
Even if one forgets about the obvious non minimality of such a point of view (the need to envisage a constant branching of worlds to describe a single `reality' experienced by communicating observers),
one can be frustrated by the fact that decoherence theory never puts into equations the apparent collapse or branching. Said differently, one can be frustrated by the fact the many world interpretation, which is just a mere interpretation of a given physical theory, and not a physical theory nor a part of a theory, exonerates physics for not modelling a simple basic fact of experience. 
Seen from the opposite angle, the price to pay for not modelling a basic fact of experience seems to be the necessity of adding an interpretation to a well-defined and experimentally confirmed physical theory.

The second point of view consists in taking what we experience at face value and admit that (i) quantum measurement involves a random aspect which is not taken into account in decoherence theory (ii) this randomness can be modelled within the realms of a physical theory. In this point of view, decoherence delivers the correct probabilities of measurement i.e. the correct averages, but does not model the randomness which prompts macroscopic quantum systems to evolve towards a particular state, as opposed to a classical superposition of states. Naturally, adopting this point of view prompts the search for a model of that randomness.

We first review the classical Langevin model of diffusion \cite{Ma85a,vK07a,DR10a} where 
the classical position- and momentum-variables of the Brownian particle obey stochastic equations. 
We then switch to quantum systems and retain the reduced density operator of the observed system as dynamical stochastic variable. Decoherence 
naturally takes into account that this variable is stochastic, but predicts only the dynamics of its statistical average, for example
through deterministic equations sometimes called master equations because of their similarities with non quantum master equations. 

In this article, we focus on the fluctuations of the reduced density operator around its statistical average predicted by decoherence and, particularly, on the fluctuations in those components which are diagonal in the so-called decoherence basis. Assuming that the interaction between the system and the environment is unbiased and using very general properties of density operators, we show that, as far as the diagonal components are concerned, quantum measurement can be rephrased as a gambler's ruin problem \cite{Ross09a} where the game is a martingale. The general optional stopping theorem for martingales \cite{Williams91a} then applies and delivers the Born rule. We also show that the noise acting on the diagonal components of the reduced density operator is necessarily non-linear and, therefore, admits a key contribution from the pure entanglement density operator usually neglected in decoherence computations. 

We then propose a simple, specific example where computations can be carried out explicitly and end the article with a thorough discussion of 
several important issues.

\section{A non quantum example}

Consider the system $S$ be a non quantum particle of mass $M$ diffusing in a dilute gas $E$ (environment) made of $N \gg 1$ non quantum particles of mass $m$ with $m \ll M$. This example is definite enough to allow for a simple discussion of all issues and the main argumentation and conclusion can be extended to other macroscopic systems in interaction with a fluid environment. 

Let $\Sigma = S \cup E$ be the total system. The motion of $S$ can be studied in various manners. The first one consists in writing the mechanical, Hamiltonian equations of motion for all particles in $\Sigma$. The natural variables at this stage are the positions and the momenta of all particles. It is well-known that the exact Hamiltonian system of equations cannot be solved analytically. One is then left with the choice between (i) solving this system numerically for relatively small, unrealistic values of $N$ (ii) using approximations leading to a physically reasonable statistical treatment of the problem. Note that option (i) requires the knowledge of initial conditions for all particles of $E$, which is of course unattainable to observers. For this reason alone, option (ii) is the one to choose. If one follows this second route, one has first to choose the variables through which the motion of $S$ will be studied. The simplest possibility consists in retaining only the position of $S$, but a better treatment, essentially due to Langevin, retains both the position and the momentum of $S$.
 The Langevin approach thus eliminates the mechanical variables of $E$ but retains all mechanical variables of $S$. Once this choice has been made, modelling the motion of $S$ comes down to writing an effective equation of motion for $S$ {\sl i.e.} to determining an effective form of the force which acts on $S$ because of its interaction with the environment. 

The Langevin models states that, at all times, the motion of the $S$ particle obeys the stochastic equations:
\begin{eqnarray}
dr_t & = & \frac{1}{m}\, dp_t \nonumber \\
dp_t & = & - \alpha p_t + \sqrt{2D} dB_t,
\label{eq:Langevin}
\end{eqnarray}
where $r_t$ and $p_t$ are respectively the position and momentum of the $S$ particle at time $t$, $\alpha$ is a positive friction coefficient and $D$ is the diffusion coefficient in momentum space.

Let us now comment on this model. As a first approximation, the interaction between $S$ and $E$ can be viewed as the result of separate interactions between $S$ and all the various particles constituting $E$. The interaction between $S$ and an $E$ particle only takes place when both are close enough and can therefore be envisaged as a collision. Thus, $S$ undergoes a certain number of collisions per unit time, which defines a mean collision time $\tau_C$. If one supposes that $E$ remains in a constant equilibrium state, each $E$ particle encountered by $S$ obeys the same statistics and this statistics is independent on the previous motion of $S$. The momentum transferred to $S$ by a collision with an $E$ particle is thus a random variable with a know statistics. This statistics depends on the statistics of $E$ and on the momentum of $S$. Now, since $m \ll M$, the relative momentum variation for $S$ in a single collision is much smaller than unity. It then takes a large number $N_c$ of collisions to substantially modified the momentum of $S$. If $m/M$ is small enough, $N_c$ is large enough and one can find an $N$ such that $1 \ll N \ll N_c$. Consider the total momentum variation of $S$ after $N$ collisions. Let $p_1, ...p_N$ be the momenta after each collision. The jump $\Delta p_i = p_{i+1} - p_i$ is a random variable whose statistics depends on $E$ and $p_i$. Since all jumps are small and the total number of
jumps is small compared to $N_c$, a reasonable first approximation is to neglect the variation of the momentum $p$ in the jump statistics. Each jump is then a random variable with 
the same statistics, which depends on $E$ and on the initial momentum $p_0 = p$. The central limit theorem then guarantees that averaging the momentum jump $\Delta p_t$ on $N$ collisions delivers a Gaussian random variable with known average and mean-square displacement. For `small enough' impulses $p_t$, a detailed computation which assumes $E$ is a dilute gas then delivers \cite{DR10a} an equation of the form
\begin{equation}
\Delta p_t  =  - \alpha p_t + \sqrt{2D} \Delta B_t,
\end{equation}
where $\Delta B_t$ is a Brownian jump. 
This equation is clearly a discrete form of equation (\ref{eq:Langevin}). The continuous equation (\ref{eq:Langevin}) is therefore only valid on time-scales much larger than $\tau_C$.
To prepare for the next section, let us remark the deterministic friction force $- \alpha p$ can be obtained by averaging the momentum jump per unit time over all realisations of the noise. This mathematical point can be rephrased using the language of statistical physics. The particle $S$ does not interact with a equilibrium statistical ensemble of copies of $E$, but with one single copy. This single copy constitutes a realisation of the noise because, for this copy, all the momenta of the $E$-particles colliding with $S$ have definite values before the collisions. The set of all these values determines a realisation of the momentum jump of the particle $S$ and, thus, a realisation of the noise acting on $S$. Averaging over the equilibrium statistical ensemble associated to $E$ comes down to averaging over all the possible values  taken before collision by the momenta of $E$-particles. By definition, the average of the noise over the statistical ensemble vanishes and one thus gets  $- \alpha p$ as the average momentum jump per unit time. It coincides with the deterministic part of that jump. The random part of the momentum jump is proportional to $dB_t$ by the central limit theorem and the proportionality coefficient codes for the mean square displacement of the momentum jump per unit time. 

Note that the Markov property of the Langevin equation is derived by assuming that $m \ll M$ is much smaller than unity, which is in turn a weak interaction assumption: it takes a time much larger than $\tau_C$ to substantially modify the momentum of $S$.

\section{Quantum Problem: General Discussion}

\subsection{The right variable}
Consider now a 
quantum system $S$ interacting with a macroscopic environment $E$.
 For example, 
the environment can be a fluid and its
constituents, conveniently called particles, be they molecules, atoms or photons. The dynamics of the system can be {\sl a priori} studied in various ways. The `mechanical' approach would 
consists in writing and solving the exact dynamical (typically Schr\"odinger) equation obeyed by the time-dependent state $\mid s^\Sigma>(t)$ of the combined system $\Sigma = S \cup E$, conceived as a system of $N$ interacting particles (degrees of freedom), $N \gg 1$.
This state defines a time-dependent `mechanical ' density operator $\rho^\Sigma (t) =  \mid s^\Sigma>(t) < s^\Sigma \mid (t)$ (see the discussion section) and the mechanical approach can be 
equivalently implemented by writing 
the exact evolution equation obeyed by the mechanical density operator $\rho^\Sigma(t)$ of $\Sigma$. Limitations similar to those discussed above for non quantum systems make it necessary to develop a statistical, effective treatment of the problem. 
%
In the non-quantum case, defining the variables retained to describe the `state' of $S$ is almost trivial. The quantum situation is more complex. As well-known, the right variable to use if one wants to implement a statistical treatment is the so-called reduced (mechanical) density operator $\rho_S(t)$ of $S$, defined by $\rho^S (t) = \mbox{Tr}_E \rho^\Sigma(t)$, where $\rho^\Sigma(t)$ is the mechanical density operator of the whole system and $\mbox{Tr}_E$ stands for a trace over the degrees of freedom of $E$. Although the ultimate aim is to develop an effective, statistical treatment of the 
dynamic of $S$ interacting with $E$, the various mechanical density operators introduced s far do not involve any averaging and, therefore, are not 
identical to the density operators commonly used in statistical physics. Indeed, the density operator of $S$ used in standard statistical physics is the average of the mechanical reduced density operator $\rho^S(t)$ over a statistical ensemble describing the statistics of the environment $E$. 

\subsection{Random process and decoherence}
In the light of what has been discussed in the previous section for the non quantum case, the right description of a single quantum system
is obtained by taking the trace over the degrees of freedom of the single environment $E$ this single quantum system is interacting with. Since this environment is macroscopic, its microscopic dynamics is best described stochastically. The reduced density operator $\rho^S (t)$ is therefore a random operator, whose law can be in principle determined once one knows the statistical properties of the environment and its interaction with $S$. 
For example, imagine a simple quantum formulation of the Langevin problem where the reduced density operator of diffusing particle corresponds to wave-packet centred on the classical, stochastic trajectory. To be consistent with the classical treatment, and because the interaction with the environment happens through collisions, the reduced density operator of the Langevin particle must be stochastic, so that the expectation values
of the position and momentum operators are indeed stochastic variables.

Keeping the discussion quite general, let us introduce the reduced density operator 
%
%
%
%
%
%
%
%
%
%
%
%
%
%
%
%
%
%
$\rho^E(t)$ of the single environment the single system $S$ is interacting with and write the density operator $\rho^\Sigma(t)$ of $\Sigma$ as
\begin{equation}
\rho^\Sigma (t) = \rho^S (t) \otimes \rho^E (t)+ \rho^e (t).
\end{equation}
Assuming all density operators are normalised to unity, one gets from the definition of $\rho^S(t)$ that $\mbox{Tr}_E \rho^e (t) = 0$. The density operator $\rho^e(t)$ is the part 
of $\rho^\Sigma(t)$ which does not factorise and, thus, describes the (time-dependent) entanglement between $S$ and $E$.

One can show \cite{Schlosshauer07a} that the most general interaction Hamiltonian $H_{\mbox int}$ fixing the time-evolution of all density operators in the interaction representation can be written as 
\begin{equation}
H_{\mbox {int}} = \sum_k H_{k}^S \otimes H_{k}^E
\end{equation}
where the operators $H_{k}^S$ ({\sl resp.} $H_{k}^E$) act only on $S$ ({\sl resp.} $E$). The evolution equation for $\Sigma$ is then
\begin{equation}
\partial_t \rho^\Sigma = - i \left[H_{\mbox {int}} , \rho^\Sigma (t) \right].
\end{equation}
Taking its trace over $E$ delivers
\begin{eqnarray}
\partial_t \rho^S (t) & = & - i \sum_k 
\mbox{Tr}_E \left(
\left[H_{k}^S , \rho^S (t) \right] \otimes
\left[H_{k}^E , \rho^E (t) \right] 
\right) \nonumber \\
& & - i \mbox{Tr}_E \left[ H_{\mbox {int}} , \rho^e (t) \right].
\label{eq:Master}
\end{eqnarray}
Note that $\mbox{Tr}_E \rho^e (t) = 0$ does {\sl not} imply that $\mbox{Tr}_E \left[ H_{\mbox {int}} , \rho^e (t) \right] = 0$. Note also that $\mbox{Tr}_E \left[ H_{\mbox {int}} , \rho^e (t) \right]$
cannot be written as the action of a linear operator acting on $\rho^S(t)$. The general equation fixing the time-evolution of $\rho^S(t)$ is thus not closed {\sl i.e.} it involves a knowledge of the entanglement between $S$ and $E$ which is not encoded in either $\rho^S(t)$ or $\rho^E(t)$ (whatever $\rho^E(t)$ may be). Thus, there is no reason to suppose {\sl a priori} that the random evolution of $\rho^S(t)$ is linear
in $\rho^S(t)$. 

Now, usual decoherence theory determines the average evolution of $\rho^S(t)$ obtained by averaging the random evolution of 
$\rho^S(t)$ over all realisations of the noise experienced by $S$. In practice, this is usually done in two steps. The first one is to replace the exact, time-dependent reduced density operator $\rho^E(t)$ of the single environment the system $S$ is interacting with by the statistical average of $\rho^E(t)$, which is typically chosen to be a Gibbs equilibrium density operator. The second step is to neglect the contribution of $\rho^e(t)$ to 
the equation of motion for $\rho^S(t)$. The success of decoherence thus suggests that the contribution of $\rho^e(t)$ to the equation of motion 
for $\rho^S(t)$ averages to $0$.

What interests us in this article is not the average equation of motion, but the fluctuations around it and, specifically, the fluctuations of the 
those components of $\rho^S(t)$ which are diagonal in the decoherence  basis. By definition, these fluctuations average to zero but must be taken into account if one wants to model the stochastic evolutions of single systems interacting with single copies of the environment. These fluctuations may come from both terms in the right-hand side of (\ref{eq:Master}). The first term generates fluctuations linear in $\rho^S(t)$. But those coming fro the second term may not be.

\subsection{Properties of the random process}
What are the properties of the random process describing the stochastic dynamics of $\rho^S(t)$? An evident one is that this random process must be trace preserving {\sl i.e.} $\mbox{Tr} \rho^S (t) = 1$ at all times. To go on, we choose to work in the so-called decoherence basis and focus on the diagonal components of $\rho^S(t)$. Each of these components is a random process. Since the random process $\rho^S(t)$ is trace-preserving, all diagonal components must be  bounded above by unity. They must also be bounded below by $0$, because each diagonal component represents the probability of finding 
the system $S$ in a particular eigenstate of the decoherence basis. 

We now suppose that the measurement induced by the interaction between $S$ and $E$ is fair or, if one prefers, unbiased, so that the noise experienced by each diagonal component of $\rho^S(t)$ averages to zero at all times. Note that this 
assumption is not only physically reasonable for many practical situations but is also implicit in usual decoherence theory, which 
describes the average effect of the noise experienced by the system $S$, and predicts that, on average, the diagonal components of 
$\rho^S(t)$ are left unchanged by the interaction between $S$ and $E$ (while the off-diagonal components of $\rho^S(t)$ naturally go to zero).

Assuming that the interaction between $S$ and $E$ is unbiased implies that the random process describing any of the diagonal components of $\rho^S(t)$ must stop when this component equals $0$ or $1$. Indeed, assume a component reaches $0$. If the noise experienced by that component has a certain non vanishing probability of shifting that component by a certain amount to the right towards positive values, it has the same probability of shifting that component by the same to the left, towards negative values. Since all components must remain positive, the noise must vanish at $0$ for all components independently. Similarly, the noise must also vanish at unity independently for all components. 

The influence of noise on quantum measurement thus comes down to a generalised
gambler's ruin problem: each eigenstate of the decoherence basis is a gambler, the corresponding diagonal component of $\rho^S(t)$ represents the fortune of the gambler at time $t$. The total fortune of all gamblers together is fixed to unity, they exchange money during the game and, when the fortune of one gambler reaches $0$, that gambler stops playing. The game goes on till one gambler wins, with fortune equal to unity. The probability that a given diagonal component of $\rho^S(t)$ reaches unity is thus the probability that the corresponding gambler wins the game. 

In mathematical terms, the random process describing the stochastic evolution of each diagonal component of $\rho^S(t)$ or, if one prefers, the fortune of each gambler, is a martingale, and the time at which that component or fortune reaches either $0$ or $1$ is a stopping time for that process. The win probability of each gambler can then be computed with the help of a so-called optional stopping theorem for martingales.

Remark that a noise with the above properties cannot be linear in $\rho^S(t)$, unless it identically vanishes. Thus, any unbiased noise acting on the diagonal components of $\rho^S(t)$ has a necessarily a contribution coming from the entanglement density operator $\rho^e(t)$ usually neglected in decoherence models.

\subsection{Optional stopping theorem and the Born rule}

There are several, slightly different versions of the optional stopping theorems. In essence, they all state that, 
under suitable, physically not stringent conditions, the expectation or average value of a martingale at a stopping time is equal to the initial expectancy or average value of that martingale. The different versions differ by the exact conditions under which they apply. For example, for a positive martingale as the ones we are dealing with in this article, Doob's version of the theorem applies, {\sl i.a.}, if the statistical average or expectancy of the martingale exists {\sl i.e.} is finite at all times and if the stopping time is bounded with probability one. We will not discuss here the differences between all versions of the theorem but refer the interested reader to 
standard references on probability \cite{Williams91a,Grimmett92a,Bhatt07a}, 
and simply assume that least one of these theorems applies to the diagonal components of the density operator $\rho^S(t)$. 
Explicit computations for a simple choice of martingale are presented in the next section as an illustration.

Since there is no ambiguity, we can drop the upper $S$ index on $\rho^S(t)$ for the remainder of this section, and we focus on 
a particular diagonal component of this operator, say $\rho^{ii} (t)$. Let $\tau$ be the stopping time for that component {\sl i.e.} the time
when that component reaches either $0$ or $1$. By the optional stopping theorem, 
\begin{equation}
< \rho^{ii}(\tau) > = < \rho^{ii}(0) >.
\end{equation}
Naturally, $< \rho^{ii}(0) > = \rho^{ii}_0$, where $\rho^{ii}_0$ is the fixed initial value of $\rho^{ii}(t)$. And, by definition,
\begin{eqnarray}
< \rho^{ii}(\tau) > & = & p^{i}_w \times 1 + p^{i}_l \times 0 \nonumber \\
& = & p^{i}_w,
\end{eqnarray}
where $p^{i}_w$ is the probability the $ii$-component `wins the game' {\sl i.e.} reaches unity before it reaches $0$, and
$p^{i}_l$ is the probability the $ii$-component `loses the game' {\sl i.e.} reaches 0 before it reaches unity. One thus finds that
\begin{equation}
p^{i}_w = \rho^{ii}_0,
\end{equation}
which is the Born rule.

\section{ A quantum example}


We now present a simple example which illustrates the above ideas and for which explicit computations can be carried out easily.

\subsection{Continuous stochastic process}

Let us consider quantum measurement in an $n$-state system $S$. The eigenstates of the measured observable are labelled by $i = 1, ..., n$. 
As in the preceding section, 
the effective density operator of the system $S$ will now be denoted by $\rho$, with components are $\rho^{ij}$, $(i, j) \in \{1, ..., n \}^2$. We normalise the trace of $\rho$ to unity and all its diagonal components, being positive, are between $0$ and $1$. What follows focuses on the diagonal components of $\rho$, the dynamics of the off-diagonal components being fully described by standard decoherence. 

The stochastic process we are presenting now is built out of $n(n-1)/2$ independent Brownian motions and codes for a 
trace preserving random walk in diagonal $\rho$ space, which randomly takes from one diagonal component to give to another. In components, the stochastic equation obeyed by the diagonal components of $\rho$ reads:
\begin{equation}
d\rho^{ii}_t =  \sigma^{ii}_K(\rho_t) dB^K_t,
\end{equation}
where summation over lower and upper repeated indices is implied (Einstein summation convention). 
The index $i$ 
runs from $1$ to $n$ and $K$ runs
from $1$ to $n (n-1)/2$.
The index $K$
can therefore be used to index the pairs $\left\{k,l\right\}$ of integers between $1$ and $n$ for which $k \ne l$. We thus replace the notations $B_K$ and $\sigma^{ii}_K(\rho_t)$ by $B_{\left\{k,l\right\}}$ and $\sigma^{ii}_{\left\{k,l\right\}}$, with the convention that the first integer appearing in the label of the pair is smaller than the second ({\sl i.e.} $k < l$). Let $D$ be a positive constant. The noise $\sigma$ is defined by 
$\sigma^{ii}_{\left\{k,l\right\}} (\rho_t) = + D$ if $i = k$ and $\rho^{kk}_t \ne 0 \ne \rho^{ll}_t$,  $\sigma^{ii}_{\left\{k,l\right\}} (\rho_t) = - D$ if $i = l$ and $\rho^{kk}_t \ne 0 \ne \rho^{ll}_t$, and $\sigma^{ii}_{\left\{k,l\right\}} (\rho_t) = 0$ otherwise. 
Thus, once a component reaches $0$, the noise coefficients coupling that component to the other ones vanish, so that component stays at $0$ until the end of the process. The process finishes when all but one component vanish. This non vanishing component then equals unity because the process is trace preserving. 

%
%
%
%
Suppose for example that $n = 3$ and the process starts from an initial density operator $\rho_0$ with no vanishing diagonal component, which is the generic situation. The equations of motion  for the diagonal components of $\rho$ initially are:
\begin{eqnarray}
d \rho^{11}_t & = & D dB^{12}_t + D dB^{13}_t \nonumber\\
d \rho^{22}_t & = & - D dB^{12}_t + D dB^{23}_t \nonumber\\
d \rho^{33}_t & = & - D dB^{13}_t - D dB^{23}_t.
\end{eqnarray}
%
%
Note that the trace of $\rho$ is conserved by this process. Note also that the opposite of a Brownian motion is a Brownian motion. Thus process therefore treats all components
of $\rho$ on equal footing.

The above equations are supposed to be valid until one of the diagonal components of $\rho$, say $\rho^{22}$ reaches $0$. The noise coefficients in front of all $dB^{12}$'s and $dB^{23}$'s then drops to zero and remain there till the end of the process, so the component $\rho^{22}$ remains also fixed at $0$ and the process goes on with the other $2$ components:
\begin{eqnarray}
d \rho^{11}_t & = &  D dB^{13}_t \nonumber\\
d \rho^{33}_t & = & - D dB^{13}_t.
\end{eqnarray}
The process stops when either $\rho^{11}$ or $\rho^{33}$ vanishes. Since the process conserves the trace of $\rho$, the non-vanishing component is then necessarily equal to unity and this signals the end of the process. The remaining component corresponds to the measurement result. 

%


\subsection{Discrete formulation as a gambling problem}

The connection with a gambling problem is best seen by discretizing the above process. We consider the time $t$ to be an integer multiple of a certain time-step $\Delta t$, 
$t = 0, \Delta t, 2 \Delta t, ..$, and suppose that the diagonal components of $\rho$ only take the discrete values $0, \Delta \rho, 2 \Delta \rho, ..., N_0 \Delta \rho = 1$. We also 
suppose as before that the initial values of the diagonal components of $\rho$, $\rho^{11}_0 = N^1_0 \Delta \rho$, $\rho^{22}_0 = N^2_0 \Delta \rho$, ... do not vanish. Note that
$\Delta \rho = 1/ (\sum_i N^i_0) = 1/N_0$ with $N_0 = \sum_i N^i_0$. To be consistent, the discretization thus requires $\Delta \rho$ to be the inverse of an integer. 

This set-up is viewed as a collection of $n$ gamblers, starting the game with fortunes $\rho^{11}_0$, $\rho^{22}_0$, ... and susceptible of increasing these fortunes by steps of $\Delta \rho$. The aim of the game is to get all the money {\sl i.e.} $1$. The game proceeds as follows. One round of the game is represented by $\Delta t$. Each round is made of 
$n(n-1)/2$ sub or partial rounds, one for each pair of gamblers. Each gambler in a pair rolls a dice once. The gambler with the highest value on the dice sees his/her fortune increase by $\Delta \rho$ while the other one sees his/her fortune decrease by $\Delta \rho$. Once a gambler has no more money, he/she stops playing and the other ones
go on with the same rule till one of them gets all the money. It is straightforward to show that this game admits the process described in the earlier section as continuous limit provided
$D = (\Delta \rho)^2/(2 \delta t)$.

\subsection{The Born rule}

Let us now compute the win probabilities of each gambler in the above game. Focus on an arbitrary gambler $i$ and put yourself in his/her shoes. To do this, use a new time $t_i$ which increases by $\delta t$ each time this gambler plays, but stays constant when this gambler does not play. If one monitors the fortune of this gambler as a function of $t_i$, one finds a simple, non-biased random walk with step $\Delta \rho$ - or, in the continuous case, a Brownian motion indexed by $t_i$. The probability this gambler wins ({\sl resp.} loses) the game is the probability this random walk - or the continuous Brownian motion- reaches $1$ ({\sl resp.} $0$). Now, computing the probability that a random walk or a Brownian motion which starts between $0$ and $1$ hits $1$ before $0$ is a standard exercise in probability. Let's present this computation for random walks.

Let $P(N^i_0)$ be the probability that a non biased random walk which starts at $N^i_0 \Delta \rho \in (0, 1)$ reaches $1$ before $0$. Evidently, $P(0) = 0$ and $P(N_0) = 1$. 
Since the random walk, at each time-step, has an equal probability $1/2$ of going to the right or to the left, one has:
\begin{equation}
P(N^i_0) = \frac{1}{2} \left( 
P(N^i_0 + 1) + P(N^i_0 - 1)
\right).
\end{equation}
This can be rewritten as
\begin{equation}
P(N^i_0 + 1) - P(N^i_0) =  P(N^i_0) - P(N^i_0 - 1).
\label{eq:lin}
\end{equation}
It follows from this that
\begin{equation}
P(N^i_0 + 1) - P(N^i_0) =  P(1) - P(0) = P(1),
\end{equation}
so that
\begin{eqnarray}
P(N^i_0 + 1) - P(1) & =  & \sum_{k = 1}^{N^i_0} \left( P(k+1) - P(k) \right) \nonumber \\
& = & \sum_{k = 1}^{N^i_0} P(1) \nonumber \\
& = & N^i_0 P(1),
\end{eqnarray}
which delivers
\begin{equation}
P(N^i_0 + 1) = (N^i_0 + 1) P(1).
\end{equation}
The win probability law is thus linear in the initial fortune.
Since $N^i_0$ is arbitrary, one can choose $N^i_0 = N_0 - 1$ in the above equation, which delivers
\begin{equation}
P(N_0) = N_0  P(1).
\end{equation}
Since $P_N = 1$, this implies $P(1) = 1/N_0$, which leads to 
\begin{eqnarray}
P(N^i_0) & = & \frac{N^i_0}{N_0} \nonumber \\
& = & N^i_0 \Delta \rho.
\end{eqnarray}
This result could have been found in a slightly shorter way by noting that (\ref{eq:lin}) combined with $P(0) = 0$ implies that $P$ is a linear function of its variable and the coefficient
can then be found by normalization or, equivalently, by using $P(N_0) = 1$.

The win probability of gambler $i$ thus coincides with his/her initial fortune. Since the fortunes actually represent the diagonal components of the density operator, this result 
coincides with the Born rule from quantum mechanics.

\section{Discussion}

We have argued that the reduced density operator of a {\sl single} quantum system interacting with a macroscopic environment is generally a stochastic, or random process and that decoherence only describes the average of this random process over all realisations of the environment. We have also argued that, because of the entanglement between the system and its environment, the dynamics 
of the reduced density operator of the observed system is generally non-linear. Thus, the time evolution of a single quantum system is best described by a possibly non-linear stochastic process in its reduced density operator. This process must be trace preserving and cannot allow any diagonal component of the density operator to become negative. Further assuming that the measurement induced by interaction with the environment is unbiased ensures the process is a martingale which stops when all components but one vanish. Unbiased quantum measurement can thus be reformulated as a gambler's ruin problem where the game is a martingale for each gambler. The Born rule then emerges as a straightforward consequence of the so-called optional stopping theorem for martingales.

Master equations are differential or finite difference equations obeyed by the time-dependent
law of a given stochastic process. In classical, non quantum statistical physics, the variables susceptible of obeying stochastic differential equations are positions and momenta and the law of a stochastic process in positions and momenta can be expressed by a function of these variables. For example, if the process is continuous, its law is actually a measure in phase space, usually represented by its density with respect to a reference measure (say, the Lebesgue measure), and  the master equation obeyed by this density is a transport equation in phase space. The equations obeyed by 
positions and momenta are stochastic but the master equation obeyed by the law of the process is naturally deterministic. 

The right variable to be used in describing the evolution of a {\sl single} quantum system $S$ interacting with its environment is the reduced density operator of that system. Since the environment is macroscopic, the interaction 
necessarily involves some randomness and it is the reduced `mechanical' density operator $\rho^S$ of the system, as the natural quantum dynamical variable, which must then obey a stochastic differential equation. The time-evolution of this reduced density operator is thus a stochastic process. The time-dependent law of that process is defined by a measure in $\rho^S$-space and the density of that measure with respect to a reference measure obeys a certain master equation. Thus, in this case, the proper master equation is {\sl not} the stochastic equation obeyed by the reduced density operator $\rho^S$, nor
its deterministic average, but the equation obeyed by the law of that operator. Nevertheless, the literature on open quantum systems and quantum noise \cite{GZ00a} often designates by `master equation' the deterministic average of the stochastic equation obeyed by the reduced density operator $\rho^S$, because of its formal similarities with master equations for non-quantum problems, and because stochastic equations for $\rho^S$ do not seem to have been considered before this article. 


As noted above, the only linear, unbiased noise acting on the diagonal components of $\rho^S$ in the decoherence basis vanishes identically. There are thus only three possibilities. 

The first possibility is the one explored in the present article: the interaction with the environment produces a non vanishing unbiased noise acting on the diagonal components of $\rho^S$. This noise is necessarily non linear. It therefore admits a contribution from the pure entanglement density operator $\rho^e$. It is also this noise which is ultimately responsible for the apparent collapse and for the Born rule. 
It is thus the entanglement {\sl i.e.} the delocalisation of correlations, which is responsible, for all aspects of quantum measurement, from decoherence to the apparent collapse and the statistics of measurement. 

The second possibility is that the interaction with the environment produces a biased noise on the diagonal components of $\rho^S$. But the bias would then be detectable as such, because changing environment would modify the statistics of measurement and because these would not depend only on the state of the quantum system before measurement.
The third possibility is that the interaction with the environment does not produce any noise on the diagonal components, and only on the off-diagonal ones, that noise averaging into standard decoherence. But if noise is present on the off-diagonal components, why would there not be noise on the diagonal ones, especially if all observations confirm a robust consequence of such a noise?

Indeed, let us stress again that the results presented in this article are valid for any unbiased noise acting on the diagonal components of the reduced density operator $\rho^S$, provided only some `light' conditions are fulfilled, as for example that the average measurement time is finite. It is not even necessary that the diagonal components a stochastic differential equation as the Langevin equation or as the dynamical equations of the example presented in Section $4$ . Thus, the ubiquity of noise explains the ubiquity of the apparent collapse associated to measurement, and the robustness of the optional stoping theorem explains the robustness of the Born rule.


Consider now two physicists observing the same quantum system $S$. Each physicist regroups the degrees of freedom of the universe to which she has no observational access into what she calls `the' environment. Suppose one of the physicist witnesses a measurement on $S$ because of the stochastic interaction of $S$ with certain degrees of freedom of what this physicist calls `the environment'. If the environment of the other physicist contains the same degrees of freedom, then this other physicist also witnesses a measurement, with the same outcome and, hence, both physicists experience the same `reality'.

%
%
%
%
%
%

Several things remain to be done. First, explore systematically simple models of quantum measurement, for example based on quantum walks, for which the stochastic evolution of $\rho^S$ could be analysed in full detail. Second, identify and perform experiments which would allow the observation of this stochastic evolution. Finally, quantum measurements in relativistic systems should be revisited in the light of the results presented in this article.

\paragraph{Acknowledgments}
The author thanks M. Brachet, Y. Ollivier and J.M. Raimond for very helpful and enlightening discussions.

\nocite{*}
\bibliographystyle{eptcs}
\bibliography{Debbasch}

\end{document}